\documentstyle[12pt,twoside]{article}
\input{epsf}
\begin{document}
\setlength{\baselineskip}{12pt}
\hoffset 0.65cm
\voffset 0.3cm
\bibliographystyle{normal}

\vspace{48pt}
\noindent
{\bf Crossover Between Flexoelectric Stripe Patterns and\\
Electroconvection in Hybrid Aligned Nematics}

\vspace{48pt}
\noindent
V.~A. DELEV$^a$, A.~P. KREKHOV$^{a,b}$ and 
L. KRAMER$^b$

\noindent
$^a$Institute of Molecule and Crystal Physics, Russian Academy of
Sciences, 450025 Ufa, Russia;
$^b$Physikalisches Institut, Universit\"at Bayreuth,
D-95440 Bayreuth, Germany

\vspace{36pt}
\noindent
We report experimental and theoretical results on the flexoelectric
instability and crossover between flexoelectric domains and
electroconvection in a hybrid aligned nematic MBBA
under d.c. voltage.
At threshold a spatially periodic flexoelectric deformation
in the form of longitudinal domains (along the planar director) 
was observed.
With increasing voltage the director deviation out of the initial plane
becomes clearly detectable.
Observations of tracers show that hydrodynamic flow develops
inside the  flexoelectric stripe pattern in the form of circular orbits 
perpendicular to the domain stripes.
With further increasing d.c. voltage electroconvection sets in
in the form of drifting oblique rolls inside the flexoelectric domains.

\vspace{24pt}

\vspace{36pt}
\baselineskip=1.4 \baselineskip

\noindent
{\bf INTRODUCTION}
\vspace{12pt}

\noindent
Liquid crystals subjected to an external electric fields
undergo various structural transformations due to the anisotropy of the
physical properties. 
A well-known instability where patterns have been studied
extensively in experiment and theory is electroconvection (EC) in 
nematic liquid crystals (NLCs).
Here one found normal or oblique rolls at onset, zig-zag, 
skewed-varicose patterns and abnormal rolls as a results of secondary 
instabilities (see, e.g., \cite{KP95} for review).

The flexoelectric effect \cite{Meyer69} can also lead to an electrically 
driven pattern-forming instability in NLCs which is, however, of a different
nature.
In particular, the flexoelectric instability in planarly aligned nematics
leads to a static periodic deformations of the director in the
form of longitudinal domains \cite{BP77,BBTU78,HV79}.

However, the role of flexoelectricity in EC
instabilities in NLCs has not been explored experimentally in much
detail.
It was found theoretically that in the case of d.c. applied voltage
for planarly aligned MBBA the
flexoelectric effect leads to a reduction of the threshold and to
the appearance of oblique or possibly parallel rolls 
\cite{MR89,TZK89,KBPTZ89} depending on the choice of the flexoelectric
coefficients.

In this paper we present experimental and theoretical results on the
flexoelectric instability in a hybrid aligned nematic (HAN) where
the director has planar alignment at one confining plate and
homeotropic at the other.
For the first time the flexoelectric pattern forming instability in
MBBA in this geometry was observed and the onset of drifting
electroconvection rolls on the background of the flexoelectric
stripes was found under d.c. voltage applied across a HAN cell.

%**********************************************************************
\vspace{24pt}
\noindent
{\bf EXPERIMENTAL}
\vspace{12pt}

\noindent
We used a standard capacitor-type cells with the 
NLC MBBA
sandwiched between two parallel glass plates with transparent electrodes
(ITO) separated by mylar spacers.
The cells of three different thicknesses $13$, $23$ and 
$40$~$\mu$m and lateral dimension $1$~cm$\times$$1.5$~cm were used.
A d.c. voltage was applied across the NLC layer.
The temperature of the cell was kept at $T=25\pm 0.1^\circ$C.

Hybrid alignment of the NLC 
was achieved by different treatment of the
confining substrates.
In order to obtain uniform planar alignment the surface of one
confining plate was rubbed in one direction.
Homeotropic alignment of the NLC at the other substrate was achieved after
cleaning of the electrode surface with alcohol.
The cells were filled from the isotropic phase of NLC and only fresh
samples were used in the experiments.

The domain patterns were observed with a polarising microscope 
Leica DMLSP 
and images were taken with a CCD-camera and digitised by frame-grabber
DT-2851
with resolution of $512 \times 512$ pixels and $256$ gray-scale levels.
The period of domain structure was obtained from the Fourier transform
of images taken at threshold.

%**********************************************************************
\vspace{24pt}
\noindent
{\bf RESULTS AND DISCUSSION}
\vspace{12pt}

\begin{figure}
\begin{center}
\vspace*{-0.2cm}
\hspace*{0.1cm}
\epsfysize=5.0cm
\epsfbox{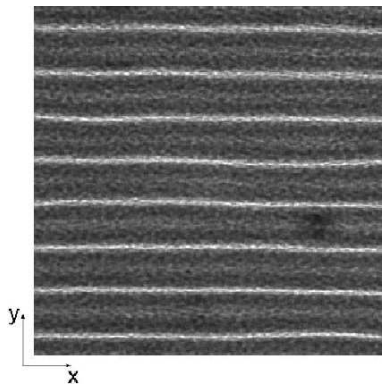}
\end{center}
\vspace*{-0.5cm}
\caption{Flexoelectric stripe pattern in hybrid aligned MBBA near the 
threshold.}
\label{fig1}
\end{figure}

\noindent
A typical flexoelectric stripe pattern in the form of longitudinal domains
(along the director orientation at the planar confining plate) 
observed slightly above some critical voltage
is shown in Fig.\ref{fig1}.
The image was taken with both polariser and analyser
parallel to the director at the lower substrate providing planar 
alignment (along the $x$ axis).
It was found that the threshold voltage for the flexoelectric domains
is independent of the polarity of the applied d.c. voltage and on
the cell thickness: $U_c=2.4 \pm 0.1$~V.
The wavelength of stripe patterns $\Lambda$ increased linearly with 
increasing thickness and $\Lambda/d=2.5 \pm 0.1$.

\begin{figure}
\begin{center}
\vspace*{-0.2cm}
\hspace*{0.1cm}
\epsfysize=5.0cm
\epsfbox{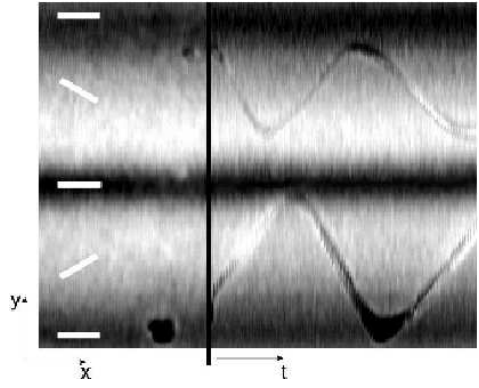}
\end{center}
\vspace*{-0.5cm}
\caption{Snapshot of one domain of flexoelectric stripe pattern 
and $y-t$ plot
showing the motion of two tracers.
The bars represent ${\bf\hat c}$ director.}
\label{fig2}
\end{figure}

At onset of the flexoelectric stripe patterns no motion of tracers
(small particles of $2-4$~$\mu$m in diameter immersed in NLC)
was detected.
With increasing d.c. voltage up to $U \approx 2.74$~V the particles
start to move slowly on circular orbits in the plane perpendicular
to the domain stripes (Fig.\ref{fig2}) signalising the development
of hydrodynamic flow in flexoelectric domains.
\begin{figure}
\begin{center}
\vspace*{-0.8cm}
\hspace*{0.1cm}
\epsfysize=5.5cm
\epsfbox{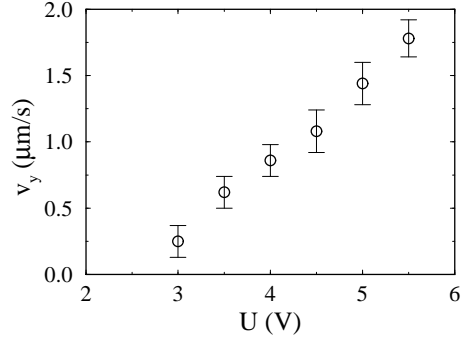}
\end{center}
\vspace*{-1.0cm}
\caption{Tracer velocity as a function of applied d.c. voltage for the
cell $d=40$~$\mu$m.}
\label{fig3}
\end{figure}
Polarised-optical analysis shows that the director deviates out of
the $x-z$ plane of the initial orientation alternatively in the
adjacent half-periods of flexoelectric domain.
Schematic distribution of the director projection onto the $x-y$ plane
(${\bf\hat c}$ director) is presented in Fig.\ref{fig2}.
The dependence of the $y$-component of the tracer velocity on the applied
voltage is shown in Fig.\ref{fig3}.

\begin{figure}
\begin{center}
\vspace*{-0.2cm}
\hspace*{0.1cm}
\epsfysize=5.0cm
\epsfbox{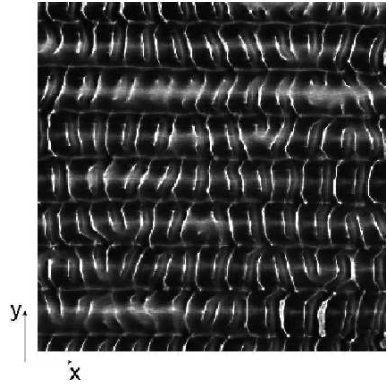}
\end{center}
\vspace*{-0.5cm}
\caption{Coexistence of stationary flexoelectric domains and drifting 
electroconvection
rolls slightly above the EC threshold.}
\label{fig4}
\end{figure}

When the applied d.c. voltage reaches the value $U_{c2}=5.8 \pm 0.1$~V
electroconvection sets in in the form of drifting rolls that break
translational invariance in the $x$ direction (Fig.\ref{fig4}).
One observes that inside of the periodic system of narrow channels formed 
by flexoelectric domains drifting EC rolls develop with some obliqueness
with respect to the $y$ axis.
The drift direction is determined by the sign of the director gradient
along the normal to the NLC layer in the initial hybrid state.
The threshold d.c. voltage, wavelength and oblique angle of EC rolls
developing on the background of flexoelectric domains are close to that
found in low-frequency a.c. electric field where the first instability
is to electroconvection (and the flexoeffect plays a small role)
\cite{ADS95,HKP95}.
With further increase of the d.c. voltage the EC mode suppresses the
flexoelectric stripe pattern and a dynamic disordered state (DSM) 
develops.
The scenario is reversible.

%**********************************************************************
\vspace{24pt}
\noindent
{\bf THEORETICAL ANALYSIS}
\vspace{12pt}

\noindent
We consider a nematic layer of thickness $d$ with 
confining plates at $z=\pm d/2$.
The director is oriented
parallel to the $x$ axis at $z=-d/2$ and parallel to the $z$ axis at
$z=d/2$.
The d.c. voltage applied across the cell is given by
$U = E_0 d$ and the electric field $\bf E$ inside the NLC layer is
${\bf E} = E_0{\bf \hat z} - \nabla \Phi$, where $\Phi$ denotes the
induced electric potential.
The equilibrium director distribution can be found by minimising the
total free energy
\begin{eqnarray}
\label{free_en}
{\cal F} = \int \{ \frac12 [ K_{11}(\nabla\cdot{\bf\hat n})^2 +
K_{22}({\bf\hat n}\cdot \nabla\times{\bf\hat n})^2 +
K_{33}({\bf\hat n}\times\nabla\times{\bf\hat n})^2 ]
\nonumber \\
- \frac12 \epsilon_0 \epsilon_a ({\bf\hat n}\cdot {\bf E})^2
- {\bf P}_{fl}\cdot{\bf E} \} dV ,
\end{eqnarray}
with the flexoelectric polarization
\begin{eqnarray}
{\bf P}_{fl} = e_{11}{\bf\hat n}(\nabla\cdot{\bf\hat n})
- e_{33}({\bf\hat n}\times(\nabla\times{\bf\hat n})) .
\end{eqnarray}
In the ground state before any instability one has a
director distribution
${\bf\hat n}_0=\{ n_{x0}(z), 0, n_{z0}(z) \}$
that is homogeneous in the $x-y$ plane.
All lengths have been measured in units of $d/\pi$ so that the confining
plates are at $z=\pm \pi/2$.
We are looking for the stability of this ground state with respect to
small director perturbations which are periodic 
in the $y$ direction ${\bf\hat n}_1(y,z) = e^{i p y} \delta{\bf\hat n}(z)$.
Then variation of (\ref{free_en}) gives the linearised equations for
$\delta{\bf\hat n}(z)$.
The solution of the linear problem was found by expanding
$\delta{\bf\hat n}(z)$ in a set of trigonometric functions
(Galerkin method, see \cite{HKP95} for detail).
The condition for existence of a nontrivial solution 
gives the neutral curve $U(p)$.
The threshold voltage $U_c$ for the flexoelectric instability
corresponds to the minimum of $U(p)$ with respect to $p$ which also
defines the critical wavenumber $p_c$.

Using the one-constant approximation 
(elastic constants $K_{11}=K_{22}=K_{33}=K$)
one obtains in the limit of small $\epsilon_0 \epsilon_a U^2/(K \pi^2)$
(then one can neglect the induced potential)
an approximate analytical expression for the neutral curve $U(p)$
\begin{eqnarray}
\label{U_p}
\tilde{U}^4 \frac{3}{2} \mu p^2 -
\tilde{U}^2 [p^2 + \mu (\frac{3}{2}p^4+2p^2+\frac12)] +
2(p^2+1)^2 = 0 ,
\end{eqnarray}
where $\mu=\epsilon_0 \epsilon_a K/e^2$, $\tilde{U}=e U/(K \pi)$
and $e=e_{11}-e_{33}$.
The threshold voltage $\tilde{U}_c$ and critical wavenumber $p_c$
obtained from (\ref{U_p}) are, in the range of validity 
$|\mu| \tilde{U}^2 \ll 1$, close to that found from the numerical
solution based on a Galerkin method.
Since the coefficients in (\ref{U_p}) do not
contain the thickness $d$ of the NLC layer explicitly, the critical voltage
and wavenumber are independent of $d$.

For the material parameters $K=6.5 \cdot 10^{-12}$~N,
$\epsilon_a=-0.53$ corresponding to the average elasticity and dielectric
anisotropy of MBBA and choosing $e=1.7 \cdot 10^{-11}$~C/m (close to
that measured recently \cite{THNUKA98}) one has
$\mu=-0.1$ and (\ref{U_p}) gives $\tilde{U}_c=2.4$, $p_c=0.96$ that
correspond to the critical voltage of flexoelectric stripe patterns
$U=2.7$~V and period $\Lambda=2.1 d$ in physical units close to that
found experimentally.

We have estimated the influence of the velocity perturbations which
were neglected in the analysis.
The development of out of $x-z$ plane director perturbations 
leads at first order in ${\bf\hat n}_1(y,z)$ to a nonzero charge density 
$\rho_{el}$.
The corresponding bulk force $\rho_{el} E_0{\bf \hat z}$ in the Navier-Stokes 
equation should be balanced by the viscous 
stresses and therefore drive a velocity ${\bf v}$.
The analysis shows that the corresponding viscous torque acting on the 
director due to the fluid motion is much smaller than the torques due 
to elastic, dielectric and flexoelectric contributions.
For the experimental value of the angle between the ${\bf\hat c}$
director and the $x$ axis, $\phi_m \approx 6^\circ$ at
$U=3.0$~V slightly above threshold, we obtain for the velocity
$v_y \approx 0.1$~$\mu$m/s in physical units
(for cell thickness $d=40$~$\mu$m)
which is of the order of that experimentally found, $0.25$~$\mu$m/s 
(Fig.\ref{fig3}).

%**********************************************************************
\vspace{24pt}
\noindent
{\bf CONCLUSION}
\vspace{12pt}

\noindent
We have investigated experimentally and theoretically the flexoelectric 
instability in MBBA with hybrid director configuration. 
The following sequence of transitions has been found
with increasing d.c. voltage:
stationary longitudinal flexoelectric domains at onset inside of which
some hydrodynamic flow develops above threshold, secondary instability
to EC with drifting rolls inside the flexoelectric stripe pattern,
and finally suppression of flexoelectric domains and transition to DSM.
It was found that the threshold 
voltage of the flexoelectric domain formation is independent of the 
thickness of the nematic layer and the spatial period of longitudinal 
domains increases 
linearly with increasing thickness which is in good agreement with 
theoretical predictions. 
For the first time we have observed the crossover between two spatially 
periodic instabilities of different nature in a single system:
spatially periodic static flexoelectric deformation of the director and
drifting EC rolls.

%**********************************************************************
\vspace{12pt}
\noindent
{\bf Acknowledgments}

\noindent
We wish to thank W. Pesch for fruitful discussions.
Two of us (V.D. and A.K.) wish to acknowledge the hospitality of the
University of Bayreuth.
Financial support from DFG (Grant 436-RUS-113/220,
Graduiertenkolleg ``Nichtlineare Spektroskopie und Dynamik'' and
Grant Kr690/14-1) and INTAS Grant 96-498
is gratefully acknowledged.
V.D. is also grateful to the INTAS Grant YSF 99-4036.

\setlength{\baselineskip}{12pt}

\end{document}